\documentclass[a4paper]{article}
\usepackage[utf8]{inputenc}
\usepackage{fullpage}
\usepackage{amsmath}
\usepackage{amssymb}
\usepackage{graphicx}
\usepackage{pdfpages}

\begin{document}



# Rate and cost of adaptation in the *Drosophila* genome

Stephan Schiffels[1], Michael Lässig[2,*] and Ville Mustonen[1,*]

[1] Wellcome Trust Sanger Institute, CB101SA Hinxton, Cambridge, United Kingdom
[2] Institute for Theoretical Physics, University of Cologne, Zülpicher Str. 77, 50937 Cologne, Germany
[*] Authors with equal contributions

Correspondence should be addressed to:
stephan.schiffels@sanger.ac.uk (Stephan Schiffels)
michael.laessig@uni-koeln.de (Michael Lässig)
vm5@sanger.ac.uk (Ville Mustonen)

## Abstract

Recent studies have consistently inferred high rates of adaptive molecular evolution between *Drosophila* species. At the same time, the *Drosophila* genome evolves under different rates of recombination, which results in partial genetic linkage between alleles at neighboring genomic loci. Here we analyze how linkage correlations affect adaptive evolution. We develop a new inference method for adaptation that takes into account the effect on an allele at a focal site caused by neighboring deleterious alleles (background selection) and by neighboring adaptive substitutions (hitchhiking). Using complete genome sequence data and fine-scale recombination maps, we infer a highly heterogeneous scenario of adaptation in *Drosophila*. In high-recombining regions, about 50% of all amino acid substitutions are adaptive, together with about 20% of all substitutions in proximal intergenic regions. In low-recombining regions, only a small fraction of the amino acid substitutions are adaptive, while hitchhiking accounts for the majority of these changes. Hitchhiking of deleterious alleles generates a substantial collateral cost of adaptation, leading to a fitness decline of about $30/2N$ per gene and per million years in the lowest-recombining regions. Our results show how recombination shapes rate and efficacy of the adaptive dynamics in eukaryotic genomes.

## Author Summary

Because recombination takes place at a limited rate, alleles at neighboring sites in a genome can remain genetically linked over evolutionary periods. In this paper, we show that evolutionary forces generated by genetic linkage have drastic consequences for the adaptive dynamics in low-recombining parts of the *Drosophila* genome. Our study is based on a new method to analyze allele frequencies that is applicable to genome data at both high and low rates of recombination. We show that genes in low-recombining regions of the

*Drosophila* genome incur a substantial cost of adaptation, because deleterious alleles get fixed more frequently than under high recombination. This cost reduces rate and power of the adaptive process. Our results suggest that the *Drosophila* genome has evolved to minimize this cost by placing genes under high adaptive pressure in high-recombining regions.

## Introduction

Genetic linkage imposes evolutionary correlations between neighboring genomic loci. Two particular effects are well known: adaptive mutations induce genetic hitchhiking of linked neutral and weakly selected variants [1-12], and deleterious mutations cause background selection on linked sites [13-20]. Both effects reduce sequence diversity, but in different ways. Background selection caused by strongly deleterious mutations leads to an unbiased removal of genetic diversity, which can be described by a reduced effective population size [17,18]. Genetic hitchhiking in selective sweeps affects common variants in a stronger way than rare variants and, hence, distorts the shape of the allele frequency spectrum [3,10,21-23]. Both effects depend on the rate of recombination, and are expected to be strong evolutionary forces in low-recombining regions of the *Drosophila* genome.

In this paper, we integrate hitchhiking and background selection into a new method to infer rates and the genomic distribution of adaptation under (partial) genetic linkage. Our model is based on diffusion theory [24,25]. Its key new variable is the *effective rate of linked selective sweeps*, which governs the hitchhiking rate of neutral and weakly selected polymorphisms and can be estimated directly from sequence data. We model linked recurrent sweeps as a Poisson process, which is often implicit in other approaches [8,10,26,27]. Our method exploits the entire allele frequency spectrum, rather than the level of diversity alone.

In *Drosophila melanogaster*, studying linkage effects has a long history. These studies are motivated by a strong correlation between the recombination rate and observed levels of diversity [28-30], which has been attributed to background selection and hitchhiking [31-36]. Our integrated inference of hitchhiking, background selection and adaptive evolution uses data from the *Drosophila melanogaster genetic reference panel (DGRP)* [29], which consists of 168 complete genome sequences from inbred lines sampled in North Carolina, USA. We also use the recently published high-resolution recombination map by Comeron and coworkers [30] to analyze polymorphism spectra as a function of the recombination rate.

We show that linkage correlations affect the adaptive process of the *Drosophila* genome in two ways. Background selection explains the broad reduction of genetic diversity with decreasing recombination rate, which is consistent with previous findings [28-30]. In addition, the low-recombining regions (which account for 21% of autosomal sequence) are marked by strongly linked selective sweeps that generate substantial hitchhiking and distort allele frequency spectra. Our integrated inference method leads to estimates of adaptive rates

also in these low-recombining regions, which had to be excluded from most previous studies due to the confounding effects of linkage[24,26,37,38]. We find a sharp drop in the rate of adaptation, compared to high-recombining regions. In addition, we estimate rate and effect of deleterious fixations due to hitchhiking, which quantify the cost of adaptation imposed by genetic linkage.

## Results

### Probabilistic evolution of neutral sequence under genetic hitchhiking and background selection

Figure 1 illustrates the evolutionary models used for our analysis. The full model of *linked adaptation* describes the evolution of a focal genomic site, which is coupled by background selection to neighboring deleterious variants and by hitchhiking to neighboring beneficial variants sweeping through the population (Figure 1a). We describe both types of linkage interactions by summary parameters, which enter an effective single-site model for the focal genomic position. Specifically, background selection caused by deleterious mutations reduces the genetic diversity at the focal site but leaves the shape of the allele frequency spectrum invariant; this effect can be described by a reduction in effective population size [13,17]. Hitchhiking in selective sweeps reduces the diversity and changes the allele frequency spectrum. We capture the effects of hitchhiking by an *effective rate of linked selective sweeps*. This parameter measures sweeps that are close enough to the focal site to affect its alleles by hitchhiking. Together, our full model has four parameters: the scaled mutation rate $\theta$, the scaled divergence time $\tau$, the scaled effective population size $\lambda$, and the effective rate of linked sweeps, $\nu$ (*Materials and Methods* and *Supplementary Text*). To quantify the importance of both kinds of linkage interactions in different parts of the *Drosophila* genome, we compare our inference from the full linked adaptation model to results from two partial models. First, the *background selection* model retains the coupling of the focal site with neighboring deleterious variants but neglects hitchhiking (Figure 1b). This model has the three independent parameters $\theta, \tau, \lambda$ and the constraint $\nu = 0$. Second, the *unlinked adaptation* model assumes that recombination is strong enough to annihilate all evolutionary effects of genetic linkage (Figure 1c). This model has the two independent parameters $\theta, \tau$ and the constraint $\nu = 0, \ \lambda = 1$.

For all three models, we derive analytic probability distributions for the allele frequencies at the focal site (*Supplementary Text*). These frequency spectra distinguish the full linked adaptation model from the background selection model and the unlinked adaptation model, which is similar to the model introduced by Mustonen and Lässig [24] (Figure 1d). Specifically, a positive rate of linked sweeps, $\nu$, removes common variants relative to rare variants, which is correctly captured by our model. A qualitatively similar depletion of the frequency spectrum was observed in rapidly evolving populations involving multiple segregating beneficial mutations [23]. The three models also have distinct effects on summary statistics of allele frequencies such as the diversity and Tajima's D, which measures the relative abundance of intermediate-frequency polymorphisms compared to low- and high-frequency variants [39].

The background selection model lowers diversity, but leaves Tajima's D invariant. In contrast, the full linked adaptation model lowers diversity and generates negative values for Tajima's D, indicating a depletion of common variants (Figure 1e).

Our approach to model background selection as a simple reduction of the effective population size is valid if linked deleterious mutations are under sufficiently strong negative selection. For background selection caused by weakly deleterious mutations, it has been shown that the effect on the spectrum is more complicated than a simple reduction in effective population size [13,14,19,20]. For the data set of this study, our simple model captures the dominant effect of background selection, which is an unbiased removal of diversity with decreasing recombination.

Using the analytical probability distributions shown in Figure 1d, we develop an inference framework to estimate the effective parameters $\theta$, $\lambda$ and $\nu$ and various other evolutionary characteristics under the different models. As data input for this inference, we use divergence data between *D. melanogaster* and *D. simulans*, together with *outgroup-directed* allele frequency distributions in the D. melanogaster population (these spectra count the number of alleles that are different from the corresponding allele in the outgroup species *D. simulans*). To test our inference framework, we use the linked adaptation model to simulate allele frequency data for varying values of $\nu$. Our inference recovers all three parameters $\theta$, $\lambda$ and $\nu$ with good accuracy (Figure 1d and Supplementary Figure 2). For later purposes, we also test whether we can infer directional selection on the focal site itself (Supplementary Figure S2a-c), with similarly accurate results.

## Quantifying linkage effects from synonymous sites in Drosophila

We study the autosomal genome sequences of 168 inbred lines of *Drosophila melanogaster*, published in the *Drosophila melanogaster Genetic Reference Panel* (DGRP)[29]. We use the *Drosophila simulans* reference genome to orient allele frequencies of segregating sites and to determine fixed differences. We first consider all synonymous sites, binned according to the local recombination rate (see Material and Methods and Supplementary Table S1). Summary statistics of this binning clearly show a strong dependency on the recombination rate (Figure 2). First, we observe a moderate increase in the rate of divergence from *Drosophila simulans* (Figure 2a) with decreasing recombination rate. Second, the diversity decreases sharply by about a factor of 8 as the recombination rate decreases, which has been reported before [28,29] (Figure 2b). Finally, for low recombination, the allele frequency spectrum deviates from the standard spectrum. This can be seen in the growing difference between two estimators of neutral diversity (Figure 2b) and further quantified by Tajima's D [39], which drops below -1.0 for zero recombining regions (Figure 2c). While a drop in diversity alone can be explained by background selection from linked *deleterious* mutations alone, a drop in Tajima's D is in this case best explained by hitchhiking with linked *beneficial* mutations.

To quantify further the change in the allele frequency data with the recombination rate, we apply our probabilistic model to each recombination bin

separately, estimating the population size $\lambda$, the mutation rate $\theta$ and the effective rate of linked drivers $\nu$. We report maximum likelihood estimates for these parameters for all recombination bins (Figure 3a-c). With decreasing recombination rate we observe clines in the parameters, which mirror the clines in the summary statistics (Figure 2). First, the mutation rate increases mildly with decreasing recombination rate, which matches the observed increasing divergence. This is very similar for the background selection model and the linked adaptation model, with slightly higher estimates from the former (Figure 3a). Reflecting the strong drop in diversity (Figure 2b), we estimate a drop in effective population size for both models (Figure 3b), with a higher population size in the linked adaptation model ($\lambda = 0.25$ versus $\lambda = 0.17$), because it explains part of the reduction in polymorphisms by hitchhiking. Finally, the linked adaptation model estimates substantial levels of the rate of linked sweeps ($\nu \sim 10$) for recombination rates below about 0.4 cM/Mb ($0.4 \times 10^{-8}$ crossovers per nucleotide per generation) (Figure 3c). This rate of linked sweeps means that every site in this region is linked to about one sweep per 80,000 years (assuming an effective haploid population size of $N_0 = 4 \times 10^6$ [40], see Supplementary Text), which is 10 times higher than the time needed for neutral variants to fix by drift.

The strong dependency of our parameter estimates on the recombination bin is clearly visible in the allele frequency spectra (Figure 3d). First, for the bin with highest recombination, the unlinked adaptation model (gray curve) fits the data very well, consistent with no strong linkage effects in that bin. The spectrum from the lowest recombination bin has a reduced level of polymorphisms, which is well explained by the drop in effective population size, captured by both the background selection and the linked adaptation model. However, this spectrum also exhibits a substantial V-shaped deviation from the background selection model prediction (blue curve), which is well captured by the linked adaptation model (red curve). The difference between the two models is also reflected in the log-likelihood score (Supplementary Figure S6), which is much larger for the linked adaptation model in regions of recombination rates below 0.4 cM/Mb.

We note that our model does not contain demographic effects, such as bottlenecks, which have been used previously to explain the synonymous site frequency spectrum in Drosophila [41]. Indeed, demographic effects on the site frequency spectrum are negligible for this data set, given the good fits of the simple unlinked adaptation model to highly recombining sites. This does not rule out that demography affects other observables, in particular haplotype structure (see also the Discussion below). We also tested whether the deviations from the neutral spectrum in low recombining regions could be caused by directional selection on synonymous sites. We find that in order to explain the depletion of intermediate frequency polymorphisms by selection, unrealistically high selection coefficients on synonymous sites are necessary (Supplementary Figure S3), clearly inconsistent with previous observations [42].

### Mixed selection model for nonsynonymous and non-coding sites

We use a mixed model for polymorphism spectra and divergence in non-synonymous and non-coding annotation categories that consists of four

components (Supplementary Text): neutral sites, weakly or moderately selected sites that contribute to rare variants, sites with adaptive substitutions (seen as fixed differences between the two species), and conserved sites under purifying selection. The component of weakly selected sites is similar to the neutral component with one additional scaled parameter $\sigma$, which denotes the selection coefficient (Supplementary Text and Supplementary Figure S2d). Together with the three neutral parameters, the full mixed model has 7 parameters: three weights to parameterize the contributions of the four components (the fourth is set by normalization), the scaled selection coefficient $\sigma$, and the three neutral parameters $\theta$, $\lambda$ and $\nu$ introduced above. We estimate these parameters in a hierarchical way: First, we fit the neutral parameters $\theta$, $\lambda$ and $\nu$ based on the synonymous sites in each bin, as already presented above. Second, we obtain maximum likelihood estimates of the other four parameters, keeping the neutral parameters fixed. To assess the impact of hitchhiking in particular, we compare our estimates of this full model to a background selection mixed model, with the constraint $\nu = 0$, and an unlinked mixed model, with $\lambda = 1, \nu = 0$.

We divide the genome into four further broad annotation categories beyond synonymous sites: Intergenic regions, Introns, untranslated regions in exons (UTR) and nonsynonymous sites (see *Material and Methods*). For each annotation category we bin all sites according to the same recombination rate bins as for synonymous sites and then perform the conditional model estimations as described above for each bin separately. Figure 4 shows the allele frequency spectrum and the model predictions for the first and the last recombination bin for each annotation category, divided by the background selection model spectrum as estimated from synonymous sites. This serves to highlight the distortions from the standard spectrum: In both high and low recombination, we see an overall reduced diversity due to selection, and for low recombination we additionally see a relative enrichment of rare variants and depletion of common variants. As can be seen, for high recombination (upper plots), the unlinked mixed model fits the data well, consistent with the results from synonymous sites. For low recombination (lower plots), the background selection mixed model is a poor explanation for the data, which clearly exhibits the V-shaped distortion observed previously. Here, the linked adaptation mixed model fits are substantially better. This is satisfying because the background selection mixed model and full mixed models have the same number of free parameters (four), since $\theta$, $\lambda$ and $\nu$ are fixed from synonymous sites, so their performance is directly comparable.

All model parameter estimates with bootstrap error estimates are summarized in Supplementary Table S2. We find that the linked adaptation mixed model performs substantially better than the background selection mixed model for recombination values below 0.4 cM/Mb, with about 16,000 units of log likelihood difference for the lowest four recombination bins in total, which is highly significant (see Supplementary Figure S6).

### The evolutionary cause of fixed differences
Because our full mixed model allows estimation of neutral, weakly selected and adaptive fractions of sites, we can specifically estimate how these fractions

contribute to fixed differences between the two species. These estimates are shown in summary for the full mixed model in Figure 5a (see Supplementary Figure S5a for a split into annotation categories and estimates from the background selection mixed model), using a more coarse-grained binning for clarity (Methods). As can be seen, below 0.4 cM/Mb most mutations are hitchhiking with selective sweeps (dark blue). This includes neutral and weakly selected variants that have reached some frequency by drift and have then been picked up by a linked sweep. Overall, the fraction of adaptive substitutions is between 10% and 20%. Figure 5b and Supplementary Figures S5b show how this fraction of adaptive substitutions is distributed across recombination values and annotation classes. We find that in high recombining regions, between 44% and 59% of nonsynonymous substitutions are adaptive, confirming previous observations in *Drosophila* [24,29]. The second highest fraction of adaptive substitutions in highly recombining regions is seen in proximal intergenic regions (20%-25%) and in UTR (15%-20%), consistent with adaptively evolving regulatory regions in the untranslated parts of exons. In intergenic regions further away from genes and in Introns, we observe only low fractions of adaptive substitutions (around and below 15%). When we look at low-recombining regions, substitutions in nonsynonymous, UTR and proximal intergenic sequence have only a small adaptive component. In the arguably less functional categories like distal intergenics and introns, we see a more erratic pattern with an actual increase up to 30% of adaptive substitutions in zero recombining regions (Supplementary Figure S5b). While this signal may indicate increased non-coding adaptation in low recombining regions, it may also be caused by technical artifacts. In particular, alignment errors between D. simulans and D. melanogaster in non-coding sequence can cause increased rates of sequence mismatches between the two species and generate a spurious signal of adaptation.

Overall, we observe a consistently smaller fraction of adaptive substitutions for low recombination with the full linked adaptation mixed model than with the background selection mixed model (Supplementary Figure S5a). This is expected since the former provides a much better fit to the neutral component (Figure 3d) than the latter, which allows the full mixed model to explain a larger portion of the spectrum using neutral sites. Only the excess of substitutions not explained by neutral substitutions are interpreted as being adaptive.

We compared our estimates of the fraction of adaptive substitutions with the generalized McDonald-Kreitman (MK) test [43], which can be corrected for the presence of weakly selected sites [29,37]. As shown in Supplementary Figure S4, this method is much more conservative than ours, in particular in low recombining regions, where the MK estimate for the adaptive fraction is estimated to be zero for all annotation categories. This is consistent with an observation made by Messer and Petrov [44], and it highlights the importance of explicitly using the allele frequency spectrum to estimate parameters such as the adaptive fraction of substitutions. The only case for which the test and our method agree quantitatively are nonsynonymous substitutions in high recombining regions, which is the case the test was originally developed for [43].

### Fitness Flux and the selective effect of adaptive substitutions

Fitness flux measures the speed of adaptation; according to the fitness flux theorem, fitness flux is generically positive [45]. In an adaptive process driven by the substitution of beneficial mutations, the fitness flux is simply the product of the substitution rate and the average selection coefficient of these changes [24,45]. In our model, we do not directly infer a selection coefficient of adaptive mutations, because any such direct inference would have a high degree of uncertainty. However, we can derive an upper bound on the strength of adaptation from its hitchhiking effects on synonymous changes. As shown in Supplementary Text, the fitness flux $\Phi$ is simply related to the rate $\nu$ of linked sweeps and the recombination rate $r$,

$$\Phi = C \, \nu \, r$$

with a proportionality constant $C$ that is greater than but of order one. Since our method cannot infer arbitrarily low levels of hitchhiking (in fact we find $\nu = 0$ for most recombination bins), we set $\nu = 1$ as a detection threshold and use the above equation to estimate an upper bound on the fitness flux based on that threshold. For nonsynonymous sites, Figure 5c shows this upper bound across the genome, except for zero-recombining regions (the lowest bin). Most estimates are between 10 and 100 in units of $\mu/2N_0$ per site, which is consistent with previous estimates in *Drosophila* [24,46]. We then use this upper bound on the fitness flux together with the inferred rate of adaptive nonsynonymous substitutions to estimate an upper bound on their selection coefficient (Supplementary Text). We find that in the part of the genome with the lower 50% of recombination rates, this upper bound on the selection coefficient is about $s_a < 1 \times 10^{-4}$, and about five times higher in regions with higher recombination, i.e. $s_a < 5 \times 10^{-4}$. Previous work in *Drosophila* has led to estimates across four orders of magnitude, $s_a = 10^{-5}$ [38], $s_a = 10^{-4}$ [24], $s_a = 10^{-3}$ [27,47], and $s_a = 10^{-2}$ [26] (see also [40] for a partial summary), some of which exceed our estimated upper bound by a factor 100.

### The cost of adaptation in low-recombining regions

We can use our model to estimate the cost of adaptation imposed by genetic linkage. This cost is given by a negative component of the fitness flux, $\Phi_-$, which is the product of the rate of deleterious substitutions, which mainly fix via hitchhiking, and their average selection coefficient, which is negative (Supplementary Text). Figure 5c shows an estimate of the resulting hitchhiking flux $\Phi_-$ for nonsynonymous sites in different recombination bins. In the lowest-recombining regions of the *Drosophila* genome, we find that $\Phi_-$ reaches values of about $1.6 \, \mu/2N_0$ per sequence site, or about $30/2N_0$ per million years per gene (Supplementary Text). This cost reaches about 5% of the upper bound on the total fitness flux in the second-lowest recombination bin (Figure 5c).

A related cost measure is the contribution of deleterious hitchhiking to genetic load [48-52]. Our mixed model predicts the stationary probability of any site to be fixed in a low-fitness allele (Supplementary Text). Multiplying this probability with the single-site selection coefficient, we obtain a genetic load of about $40/2N_0$ per gene. This load measures the fitness cost of placing an average gene

into a region of low recombination; its evolutionary interpretation is discussed below.

## Discussion

In this study, we have developed an analytic model for adaptive evolution under partial genetic linkage. This model maps the complex process of correlated multi-site evolution onto an effective single-site process with three evolutionary forces: positive selection causing primary adaptation, genetic draft inducing hitchhiking, and background selection constraining diversity and divergence. Despite its simplicity, our model explains allele frequency data across different recombination classes of the *Drosophila* genome with remarkable accuracy (Figures 3d and 4). Because our inference method does not use haplotypes, it can be applied to bulk sequencing data, which extends its possible range of applications.

Consistent with previous studies, we infer high rates of adaptive evolution in high-recombining sequence of the *Drosophila* genome: about 50% of the nonsynonymous substitutions in coding sequence, and 20% of the substitutions in UTR and in proximal intergenic sequence are adaptive. We obtain upper bounds for the resulting speed of adaptation, which is measured by a fitness flux of order 100 $\mu/2N_0$ per sequence site, and for the average selection coefficient of adaptive changes, which is of order $10^{-4}$. These bounds follow from a simple argument: adaptive processes with higher total fitness flux would distort the frequency spectrum of synonymous polymorphisms, which we do not observe in the high-recombining regions spanning 80% of the *Drosophila* genome. This argument constrains the *average* speed of adaptation by *hard* selective sweeps, which lead to substitutions of the beneficial allele and drive the long-term adaptive divergence between species. It does not exclude individual selective sweeps with far higher selection coefficients. It also does not constrain soft and partial sweeps, which involve beneficial alleles arising on diverse genetic backgrounds or alleles with a conditional selective advantage [53]. These sweeps leave a weaker trace in the synonymous frequency spectrum that hard sweeps. Soft sweeps have been inferred by haplotype-based genomic scans for adaptation in several systems including *Drosophila* [54].

Remarkably, the allele frequency spectra of the North Carolina flies lack a clear footprint of the population's recent demography. About 80% of synonymous sites in the autosomal genome show a textbook neutral spectrum with constant effective population size. The spectrum in the 20% lowest-recombining sequence sites is depleted of common variants, but we attribute this recombination class-specific signal to hitchhiking rather than to recent changes of population size. We emphasize that our analysis is based on site frequency spectra, so this result does not rule out demography being visible in some other (e.g., haplotype-based) observables. In other systems, for example in humans, demographic effects are more prevalent in the allele frequency data, and our method will have to be extended to distinguish them from signals of adaptation and of linkage correlations. Similarly, we can extend our method to account for a variable density of functional elements, say gene content. This is expected to

generate heterogeneous amounts of adaptation, hitchhiking, and background selection within one recombination class.

The most striking result of this study is a strong quantitative relation between the amount of adaptation and linkage correlations in the *Drosophila* genome, which is summarized in Figure 5. The fraction of adaptive amino acid substitutions drops from about 50% in high-recombining regions to small values in the 20% lowest-recombining sites; a similar drop is observed in UTR and proximal intergenic regions. The majority of substitutions in all of these sequence classes can be accounted for by hitchhiking. We also have shown that hitchhiking imposes a substantial cost on adaptation, which is measured by a negative fitness flux component $\Phi_-$ of about $30/2N_0$ per million years per gene and a genetic load of about $40/2N_0$ per gene. Together, we obtain a complex picture of adaptation in low-recombining regions: linkage interactions reduce rate and power of *primary* selective sweeps by hitchhiking. In a continual adaptive process, the fitness cost of hitchhiking is compensated by a cascade of *secondary* adaptive changes at the hitchhiking sites. This complexity of the genomic dynamics of adaptation is a generic consequence of linkage interactions, which become a strong evolutionary force under low recombination [11,55-57]. We have shown that the interplay of adaptation and linkage interactions already generates strong effects in *Drosophila*, a species with overall high recombination rates. These effects are expected to be even stronger in other species with lower recombination rates that result, for example, from alternating sexual and asexual reproductive modes. The salient point about *Drosophila* is that recombination rates and, hence, the strength of linkage correlations vary strongly within its genome, with a broad decrease from central to distal parts of the chromosomes [30]. Thus, our findings may suggest that the distribution of genes in the *Drosophila* genome results, in part, from an adaptive minimization of the cost of adaptation: genes under high adaptive pressure are predominantly placed in high-recombining genomic regions. In this way, the interplay of adaptation and genetic linkage can shape the large-scale genome architecture.

## Material and Methods

### Genomic Data

We downloaded the complete genome sequences of 168 lines from the Drosophila Melanogaster Reference Panel (DGRP) from the DGRP website (http://dgrp.gnets.ncsu.edu) as fasta files. We downloaded the reference sequence from *Drosophila simulans*, aligned to the reference sequence of *Drosophila melanogaster*. We computed outgroup directed allele frequencies at all sites at which a) there is a valid *Drosophila simulans* allele, b) at least 150 lines of the DGRP sequences have a called allele (see Supplementary Table S1). To simplify downstream analysis, we normalized all sites to 150 called alleles. Specifically, if $m \geq 150$ alleles are called, and $k$ of those are different from the simulans allele, we computed the normalized outgroup directed allele frequency (allele count) as

$$\tilde{k} = \left\lfloor 150 \, \frac{k}{m} + \frac{1}{2} \right\rfloor.$$

## Sequence Annotation

We downloaded gene annotations from flybase [58] and defined annotation categories as follows:

- INTERGENIC FAR: Intergenic regions that are at least 5kb away from genes
- INTERGENIC MEDIUM: Intergenic regions within 5kb distance to the next gene, but further away than 500bp
- INTERGENIC NEAR: Intergenic sites within 500bp of a gene
- INTRON: Introns on protein-coding genes
- UTR: untranslated regions on the exons
- SYNONYMOUS: protein-coding sites on the reference at which none of the three possible point mutation changes the encoded amino acid
- NONSYNONYMOUS: protein-coding sites on the reference at which any of the three possible point mutations changes the encoded amino acid.

In some figures we joined the intergenic categories where appropriate. Most genes have multiple associated transcripts due to alternative splicing. We chose the transcript corresponding to the longest encoded protein coding sequence for each gene and annotated introns, UTRs, synonymous and nonsynonymous sites according to this one transcript. See Supplementary Table S1 for the number of sites in a given annotation category on the different chromosomes.

## Recombination Rate Binning

Recombination maps were obtained from Comeron et al. [30] through their website http://www.recombinome.com, defined as mean rates within 100kb windows. We used the recombination map to annotate every site in the *Drosophila* genome. We then used only sites in the SYNONYMOUS annotation category on the autosomal chromosomes (2L, 2R, 3L and 3R) and defined quantile boundaries on this set. Specifically, we sorted all recombination rate values of this set of sites and determined recombination rate boundaries by dividing the data set into 21 equally large subsets of values. We then used these quantile boundaries to bin all sites (not just those in category SYNONYMOUS) into bins according to their local recombination rate. Here are the quantile boundaries used in this study for autosomal data (in cM/Mb):  0.0, 0.069, 0.217, 0.415, 0.44, 0.821, 1.055, 1.29, 1.415, 1.592, 1.741, 1.938, 2.169, 2.354, 2.612, 2.838, 3.156, 3.461, 3.796, 4.244, 5.395, Infinity. For figure 5, we used a more coarse binning, merging bins [0], [1, 2], [3, 4, 5], [6, 7, 8, 9], [10, 11, 12, 13, 14, 15, 16] and [17, 18, 19, 20].

For plotting purpose only, we show allele frequency spectra with averaged number of counts in neighboring allele frequencies. Specifically, of the 151 allele frequency values (incl. 0 and 150), we average values in groups of 2 from frequency 10 through frequency 11, and in groups of 3 from frequency 12 through 44 and from frequency 120 through 149. We average values in groups of 5 from frequency 45 through 119.

## Summary of model parameter estimation

In the *Supplementary Text* we derive the allele frequency distribution for several basic models, which all derive from the full probability $P(k; m, \tau, \theta, \sigma, \nu, \lambda)$ to observe in $m$ samples of the ingroup species $k$ alleles which differ from some outgroup. The parameters are the scaled time to the common ancestor of the two species $\tau = t/2N_0$, the scaled mutation rate $\theta = 2N_0\mu$, the scaled selection coefficient $\sigma = 2N_0 s$, the scaled rate of linked selective sweeps $\nu = 2N_0 V$ and the scaled effective population size $\lambda = N/N_0$. The basic models are:

$P_{\text{unlinked}}(k; m, \tau, \theta) = P(k; m, \tau, \theta, 0, 0, 1)$,
$P_{\text{BGS}}(k; m, \tau, \theta, \lambda) = P(k; m, \tau, \theta, 0, 0, \lambda)$,
$P_{\text{linked}}(k; m, \tau, \theta, \nu, \lambda) = P(k; m, \tau, \theta, 0, \nu, \lambda)$,
$P_{\text{selection}}(k; m, \tau, \theta, \sigma, \lambda) = P(k; m, \tau, \theta, \sigma, 0, \lambda)$.

These models are used to estimate parameters based on synonymous sites only. To model other annotation categories, we introduce mixed models with the following components:

- A neutral component with fraction $c_n$, modeled by one of $P_{\text{unlinked}}$, $P_{\text{BGS}}$ or $P_{\text{linked}}$.
- A weakly selected component with fraction $c_w$, which is modeled by the most general model $P$, but with a constraint $\sigma > 1$.
- A fraction of adaptive substitutions with fraction $c_a$.
- A fraction of additional conserved sites with fraction $c_c = (1 - c_n - c_w - c_a)$.

Taking the linked adaptation model as the neutral model, the full mixed model is:
$P_{\text{linked mixed}}(k; m, \tau, \theta, \sigma, \nu, \lambda, c_n, c_w, c_a)$
$\quad = c_n P_{\text{linked}}(k; m, \tau, \theta, \lambda) + c_w P(k; m, \tau, \theta, \sigma, \nu, \lambda) + c_a \delta_{k,m} + c_c \delta_{k,0}$

Similar mixed models are defined using the background selection or the unlinked adaptation model as neutral component, with accordingly fewer neutral parameters.

To estimate parameters, we consider a data set of outgroup-directed allele frequencies with a fixed sample size $m$. We denote the number of sites with allele frequency $k$ by $n_k$. The total log-likelihood of the data given parameters $\Theta$ is then:

$$\mathcal{L}(\{n_k\}; m, \Theta) = \sum_{k=0}^{m} n_k \log(P(k; m, \Theta))$$

where $P$ is a placeholder for the appropriate model, and $\Theta$ denotes the set of model parameters. Parameter estimates are obtained by maximization of the log-Likelihood:
$$\hat\Theta = \underset{\Theta}{\text{argmax}}\, \mathcal{L}(\{n_k\}; m, \Theta).$$

As detailed in Supplementary Text, we do not simultaneously estimate all 7 parameters of the mixed model, but use a hierarchical approach, first estimating the neutral parameters from synonymous sites only.

## Bootstrapping

We use bootstrapping to obtain error estimates for all parameters and derived estimates (i.e. fractions of substitutions and genetic load). Each bootstrap sample is generated from the frequency count data in a given bin, by resampling all

frequency counts with replacement from the original counts. We obtain a standard error estimate for each parameter by taking the standard deviation of that parameter across 20 bootstrap samples.

### Simulations

We simulated the background selection and linked adaptation models using a Monte Carlo method. In the standard simulation, a population with $N = 1000$ individuals is simulated at a single site with two alleles. Mutations occur randomly with rate $\mu$. Each generation is sampled with replacement from the previous generation. We introduce selection by assigning a modified sampling weight $p = \exp(s)$, where $s$ is the selection coefficient. Linked selective sweeps occur with rate $V$. For each sweep we choose a linked allele with a probability equal to its frequency $x$. A sweep instantaneously fixes that allele, setting $x = 1$.

For a single sample, we start with an equilibrated allele frequency as ancestral value (obtained by simulating a single site for $2/\mu$ generations) and then simulate two separate populations for $t$ generations, starting with the ancestral allele frequency. We sample a single outgroup allele and 20 ingroup alleles from the two evolved populations, respectively. The result is a single outgroup-directed allele frequency $k$, obtained by counting the number of ingroup alleles that are different from the outgroup. We simulate 40,000 independent samples before testing parameter inference based on the resulting spectrum.

### Implementation

The implementation of the inference method is available under `https://github.com/stschiff/hfit`.

## Acknowledgements

# Figures

**Figure 1: Models of adaptation under linkage**
(a) *Linked adaptation model:* The evolution of a neutral or weakly selected focal site (black) involves linkage-generated interactions with neighboring beneficial mutations (red) and deleterious mutations (blue). We describe these interactions by two effective model components: deleterious mutations lower the effective population size via background selection ($\lambda < 1$), beneficial mutations generate an effective rate of linked sweeps ($\nu > 0$.) We compare this model with two partial models: (b) *Background selection model*: We include linkage interactions only with neighboring deleterious mutations and disregard hitchhiking (i.e., $\nu = 0$.) (c) *Unlinked adaptation model*: In this single-site model, the focal site evolves independently of its genomic neighborhood (i.e., $\lambda = 1, \nu = 0$.) (d) Frequency distribution of single-nucleotide polymorphisms at the focal site for the linked adaptation model (red), the background selection model (blue), and the unlinked adaptation model (gray). Analytical spectra given by our model are compared to simulations for a Wright-Fisher population (see *Material and Methods*). (e) Linkage effects on the polymorphism spectrum can be observed in the sequence diversity, $\pi$, and in Tajimas *D*, which measures the depletion of intermediate-frequency polymorphisms.

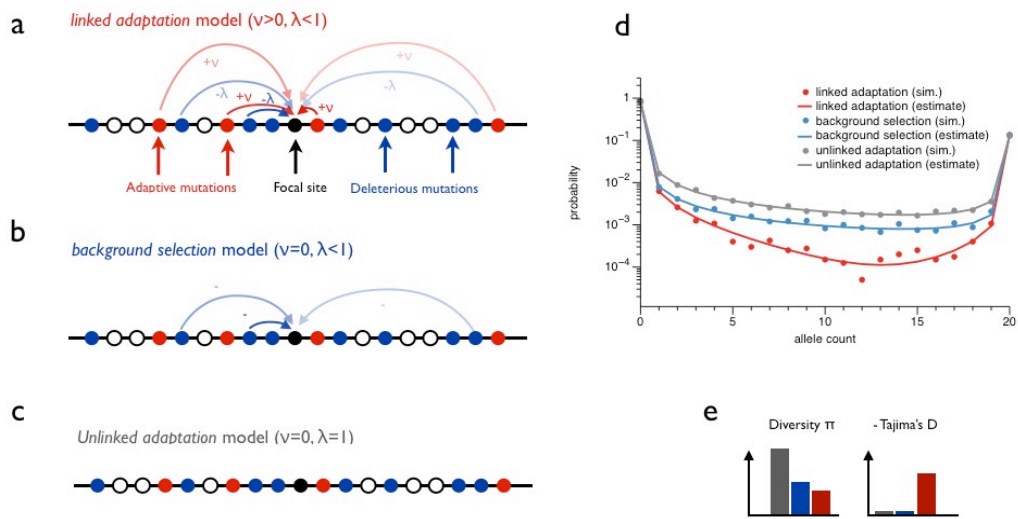

**Figure 2: Summary statistics of synonymous sites**

This figure shows summary statistics of the polymorphism and divergence data from synonymous sites in *Drosophila*, as a function of the recombination rate. The divergence (a) stays roughly constant, with a mild increase with decreasing recombination rate. The diversity (b) drops sharply as the recombination rate goes to zero, which is seen in both standard estimators $\theta_S$ and $\theta_\pi$ (see Supplementary Text). c) Tajima's D, a measure of the distortion of the allele frequency spectrum, is the normalized difference between the two standard estimators in (b). It becomes substantially negative for decreasing recombination, indicating hitchhiking.

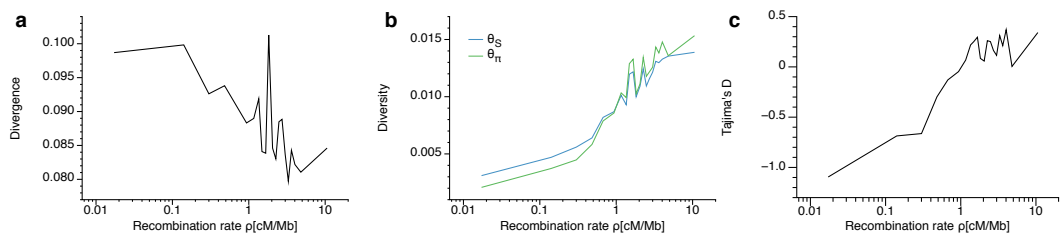

**Figure 3: Neutral model parameters from synonymous sites**

This figure shows the estimated model parameters for the linked adaptation and the background selection model from synonymous sites. Error bars are obtained by bootstrapping from the data, as detailed in Methods. With decreasing recombination rate, the mutation rate (a) increases, and the effective population size (b) decreases, but less so for the linked adaptation model (red). The effective rate of linked sweeps (c) increases sharply in regions of $\rho < 0.4 cM/Mb$. (d) Synonymous sites allele frequency spectra in low and high recombining regions in *Drosophila melanogaster* differ substantially: High recombining regions (empty circles), very closely match the expected neutral allele frequency spectrum without linkage (gray line). In low recombining regions (filled circles), the linked models (blue and red) both capture the decrease in the level of polymorphisms, but only the linked adaptation model (red) also captures the observed distortion. In (d) data points are averaged over neighboring frequency values for plotting purpose only (see Methods).

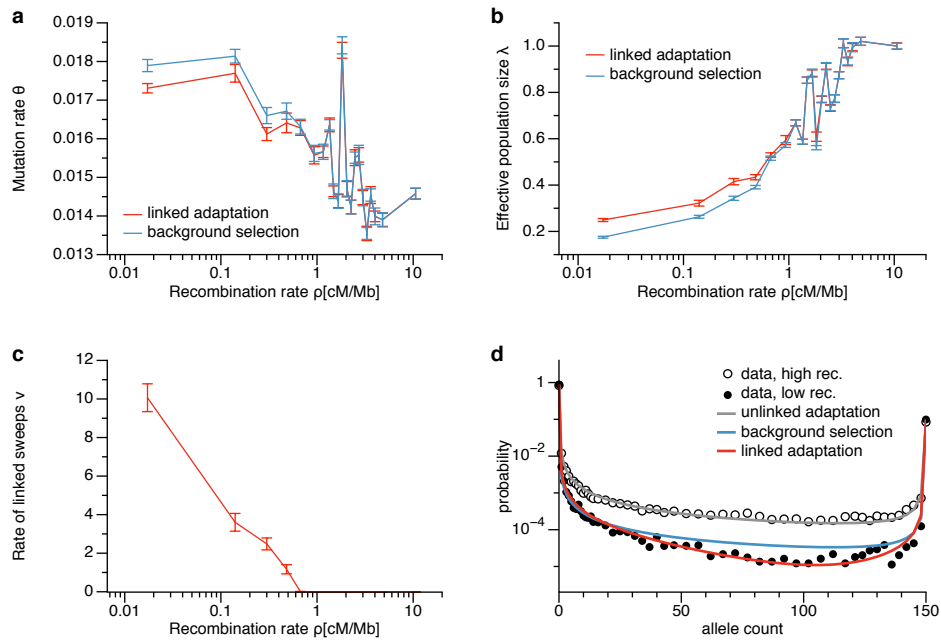

### Figure 4: Nonsynonymous and Noncoding model fits

Shown are the site frequency spectra for all non-synonymous and non-coding annotation categories, divided by the best neutral unlinked model on synonymous sites (see Figure 3d). The upper plots show data and model fit for the highest recombination bin in all annotation categories. The lower plot shows data and fits for the lowest recombination bin, for which rare variants are relatively more frequent than common variants, due to hitchhiking, as correctly captured by our linked adaptation mixed model. Data points are averaged over multiple counts for plotting purpose only (Methods).

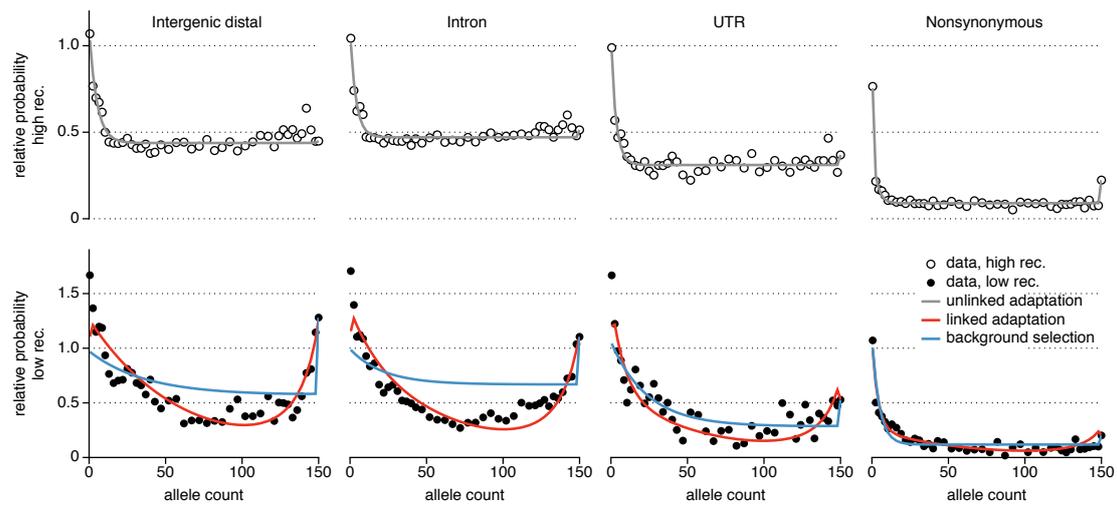

**Figure 5: Sequence divergence statistics**

This figure shows how fixed differences between *D. melanogaster* and *D. simulans* are distributed in different sequence annotation classes and for different recombination rates . (a) Overall numbers of substitutions. In high-recombining regions (with recombination rates > 0.4 cM/Mb), most substitutions are caused by genetic drift or are adaptive, while hitchhiking is negligible. In low-recombining regions (with recombination rates < 0.4 cM/Mb), hitchhiking influences most substitutions. We also observe an increase of adaptive substitutions in intergenic regions (see text). (b) Fraction of adaptive substitutions in different annotation classes. In nonsynonymous, UTR, and proximal intergenic sequence, this fraction decreases with decreasing recombination rate (cf. Figure S5 b for other annotation classes). (c) Fitness flux. Total fitness flux (upper bound, see text) and negative (hitchhiking) component in coding sequence. These estimates indicate that the overall speed of adaptation decreases, while the cost of adaptation increases with decreasing recombination rate (see text). For clarity, In (b) and (c) we show zero valued data points with a small positive offset.

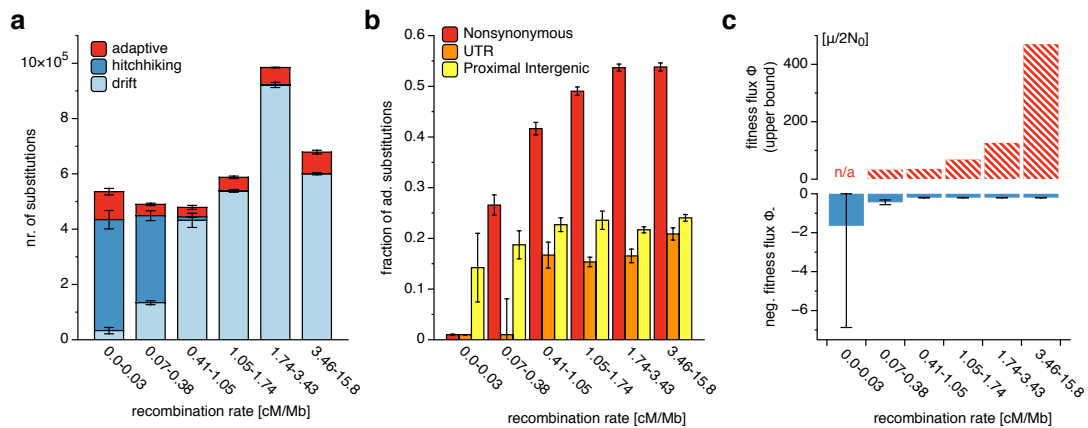

## Supplementary Figure S1: Validation on simulated data

We tested our effective model on simulated data from a Wright-Fisher population, using a forward-in-time simulation scheme (see Supplementary Text). We varied the effective rate of sweeps, *V*, keeping the other two parameters *N* and *μ* fixed, and inferred all three parameters. As can be seen, this inference works well, in particular with respect to estimating *V* (a), with some underestimation for large values. The mutation rate is slightly underestimated by our model (b), and the effective population size estimates show some degree of anti-correlation with increasing *V* (c).

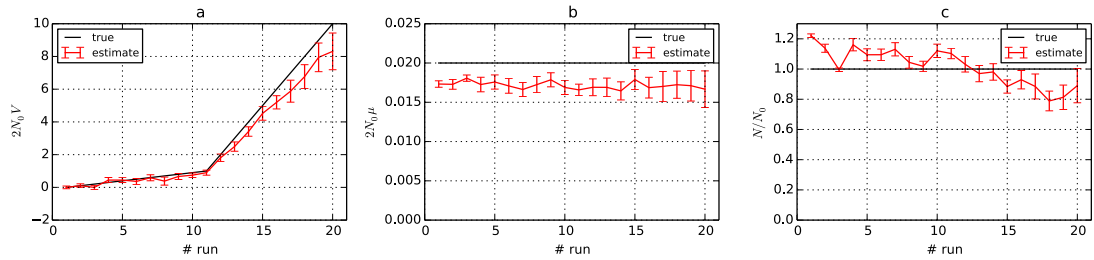

## Supplementary Figure S2: Estimating selection from simulations

(a-c) We also tested the unlinked model under selection using simulations. Here we simulated data sets varying the selection coefficient, with no linked sweeps, $V = 0$. As can be seen, the inference of all three parameters $s$, $N$ and $\mu$ works very well up to about $2N_0 s = 10$, and gets unreliable beyond that value. d) We also tested the inference of the selection coefficient with and without linked sweeps if all three neutral parameters $V$, $N$ and $\mu$ are fixed. This is relevant to our estimates from non neutral annotation classes, where we fixed these neutral parameters from synonymous sites.

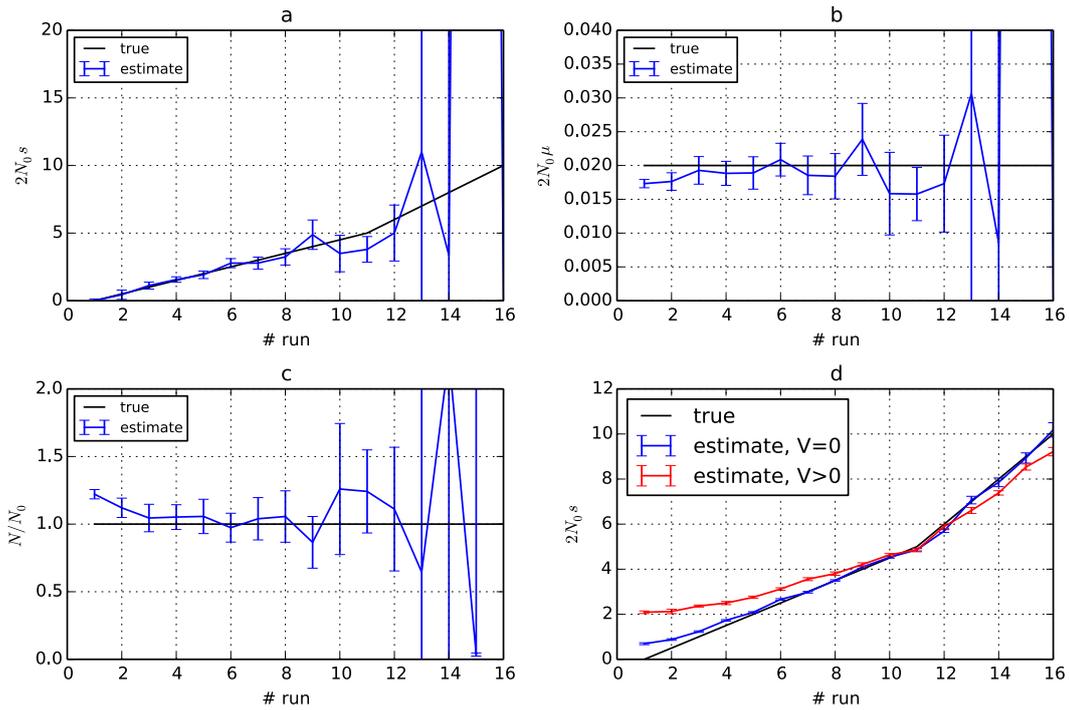

## Supplementary Figure S3: Fitting a model with selection to synonymous sites

We tested a model without hitchhiking but selection on synonymous sites (see Supplementary Text). (a-c) Inferred parameters are highly unrealistic for that model, yielding a 10 times higher mutation rate and a 10 times lower effective population size than inferred from the background selection and linked adaptation model. Most importantly, inferred selection coefficients are unrealistically high. d) With these extreme parameters, the resulting probability for allele frequencies under selection (green) is in principle able to explain to some extent the distortion of the frequency spectrum in low recombining regions, falling in between the background selection and the linked adaptation model.

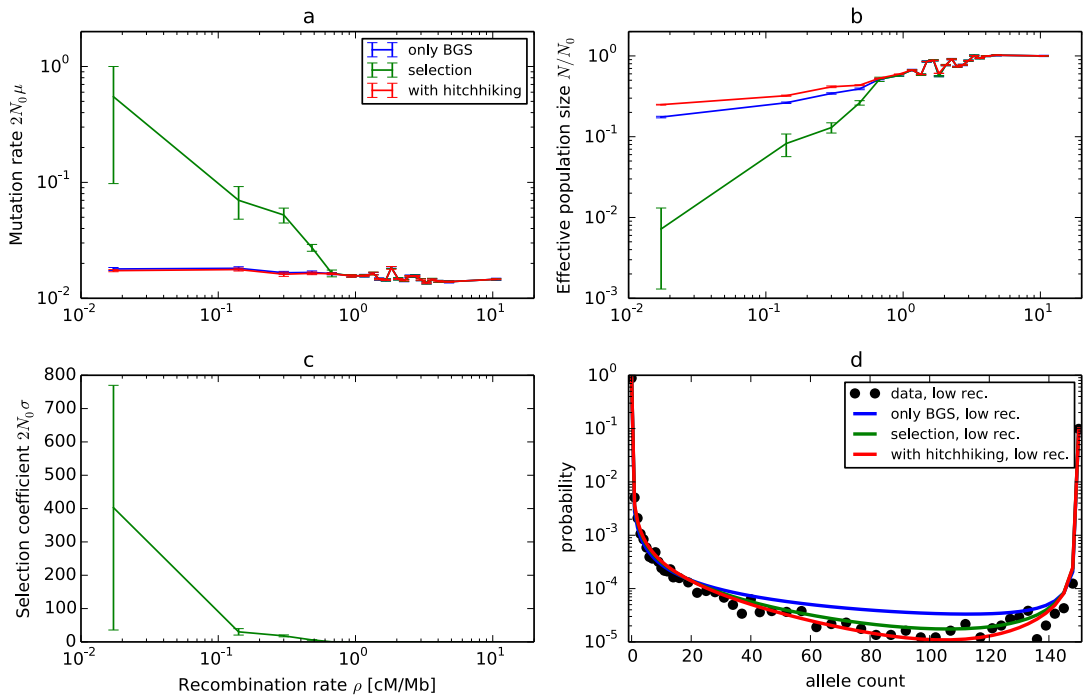

## Supplementary Figure S4: Comparison with the generalized MK-test

This figure shows a comparison of our estimates of the fraction of adaptive substitutions with estimates from the generalized McDonald-Kreitman test (MK test). For all annotation categories, we plot the predicted fraction of adaptive substitutions from the background selection (blue) and linked adaptation model (red), together with estimates from the generalized MK test as a function of the recombination bin index.

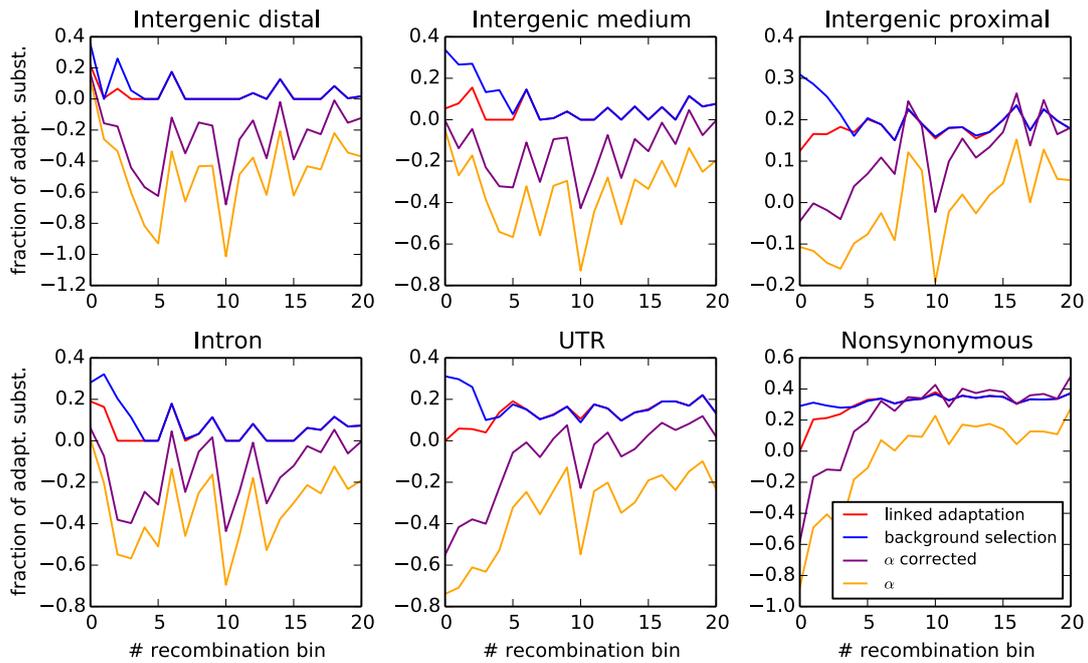

# Supplementary Figure S5: Causes of substitutions in different annotation categories

In (a) we show the data from Figure 5a, but stratified across different functional categories. (b) shows the analogous version of Figure 5b in the main text, but for the less functional categories.

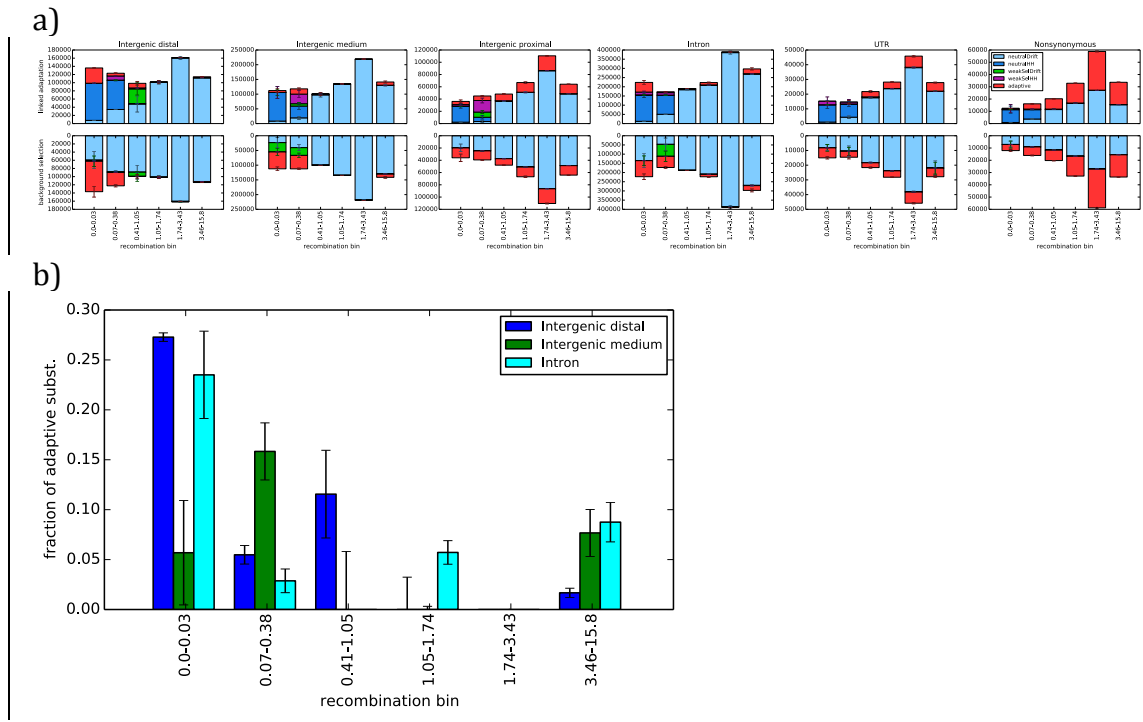

## Supplementary Figure S6: Log Likelihood difference of the linked adaptation vs. background selection model

This figure shows the difference in log likelihood between the linked adaptation and the background selection model. In most bins, the two models perform equally, but we find that in the lowest four bins, the linked adaptation model performs substantially better, with a total score difference of $14{,}149 \pm 6{,}914$, where the error estimate is obtained from bootstrapping.

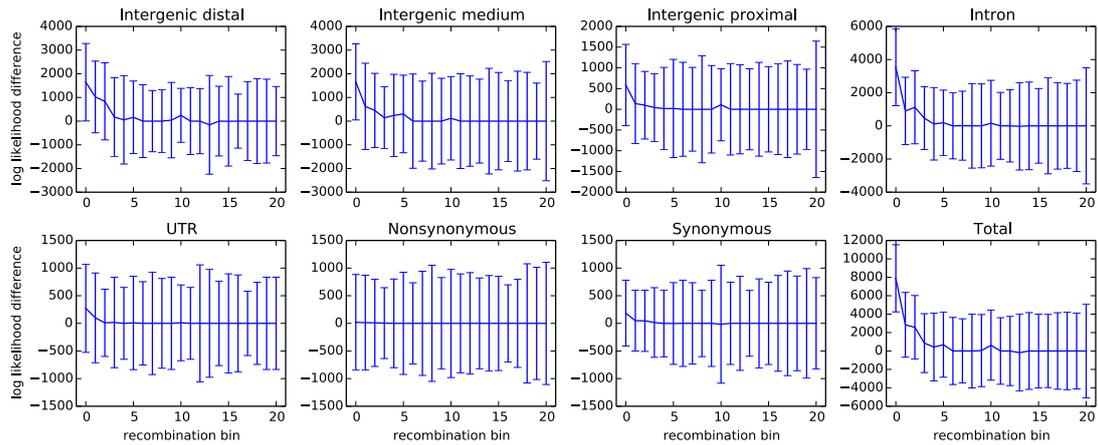



\title{Supplementary Text}
\author{}
\date{}
\maketitle

  \tableofcontents

    \section{Definitions}
    \label{sec_def}
    
    We introduce these scaled variables:
    \begin{itemize}
        \item the scaled mutation rate $\theta=2 N_0 \mu$
        \item the scaled sweep rate $\nu=2N_0 V$
        \item the scaled selection coefficient $\sigma=2Ns$
        \item the scaled time to the common ancestor of two speces $\tau=t/2N_0$
        \item the scaled effective population size $\lambda=N/N_0$. For most of the derivation, we consider $\lambda=1$ and will relax this further below.
    \end{itemize}
    
    \section{General allele frequency distribution}
    \subsection{Neutral equilibrium allele frequency distribution}
    
    We consider a single biallelic site with two alleles 0 and 1. We denote the frequency of allele 1 in the population with x. In the absence of any selective advantage, a symmetric mutational process with scaled rate $\theta$ and an effective population size $N$, genetic drift will lead to an equilibrium distribution \cite{Sawyer:1992vb} which for small mutation rates can be decomposed into two terms \cite{Mustonen:2007im}:
    \[
        Q(x;\theta)=\frac{1}{2} Q_0(\theta)+\frac{1}{2} Q_1(\theta)
    \]
    with the partial distribution
    \[
        Q_a(x;\theta) = \frac{2}{Z_a (\theta)} ((1-a)(1-x)+a x)\left(x(1-x)\right)^{-1+\theta}
    \]
    and the normalization factor
    \[
        Z_a(\theta)=\frac{\Gamma(\theta)^2}{\Gamma(2\theta)}.
    \]

    To derive the probability to observe a discrete allele count rather than a continuous allele frequency we model the binomial sampling process explicitly. The probability to observe k out of m individuals carrying allele 1 is then given as a binomial moment of the equilibrium distribution:
    \begin{equation}
        \label{eq_binom_sampling}
        M(k;m,\theta) = \binom{m}{k} \int_0^1 x^k (1-x)^(m-k) Q(x;\theta)\mathrm{d}x
    \end{equation}
    which leads again to the independent partial distributions
    \[
        M_a (k;m,\theta) = \binom{m}{k} \frac{2}{Z_a(\theta)} ((1-a)(m-k)+a k+2\theta) 
        \frac{\Gamma(k+\theta)\Gamma(-k+m+\theta)}{\Gamma(1+m+2\theta)}
    \]
    for $a={0,1}$.
    
    \subsection{Hitchhiking with recurrent selective sweeps}
    We model recurrent selective sweeps as a Poisson process with scaled rate $\nu$. The probability that no sweep occurs for a time $t$ (in scaled units of $2N$ generations) is then exponential:
    \[
        \mathrm{Prob(no sweep)}(t)=e^{-\nu t}
    \]
    
    We approximate the expected scaled time it takes for a neutral polymorphism to reach allele frequency $x$ by $t=x$ (again in units of $2N$ generations) and express the partial equilibrium probability under recurrent sweeps approximately as
    \begin{equation}
        \label{eq_qdist_hh}
        Q_a(x;\theta,\nu) = Q_a(x;\theta) e^{-\nu((1-a)x+a(1-x))} + \sum_{b=\{0,1\}}\delta(x-b)h_a(b,\theta,\nu)
    \end{equation}
    where the sum over the two delta distributions accounts for complete hitchhiking to the two fixed states $b=\{0,1\}$ and which will be given further below.
    
    We can again derive binomial moments to express the discrete probability distribution for sampling $k$ out of $m$ individuals with allele 1:
    \[
        \begin{split}
            M_a(k;m, \theta, \nu) = & \binom{m}{k}\frac{2}{Z_a(\theta)}\left(a e^{-\nu}+(1-a)\right)
            \frac{\Gamma(k+\theta)\Gamma(-k+m+\theta)}{\Gamma(m+2\theta)}\times\\
            & \left(
                a {}_1F_1(k+\theta, m+2\theta, (2a-1)) - \frac{m-k+\theta}{m+2\theta}
                {}_1F_1(k+\theta, 1+m+2\theta, (2a-1)\nu)
            \right)\\
            & + \sum_{b=\{0,1\}}\delta_{k,bm}h_a(b,\theta,\nu)
        \end{split}
    \]
    where the last term again account for the fraction of hitchhiking alleles and affects the two boundary states $k=0$ and $k=m$. Here, ${}_1F_1(a,b,x)$ denotes the confluent hypergeometric function.
    
    The hitchhiking fraction is derived by integrating the probability to hitchhike from frequency $x$ to frequency $b=0$ or $b=1$:
    \[
        h_a(b,\theta,\nu) = \int_0^1 Q_a(x;\theta) (b x + (1-b)(1-x))\left(1-e^{-\nu((1-a)x+a(1-x))}\right)\mathrm{d}x
    \]
    where the term $bx+(1-b)(1-x)$ just accounts for the fact that the allele hichhikes to $b=1$ with probability $x$ and to $b=0$ with probability $1-x$.
    Using the above defined moments $M_a(k;m,\theta)$ we can compute this integral to:
    \[
        \begin{split}
            h_a(b,\theta,\nu) = & M_a(b;1,\theta)-\frac{-2a e^{-\nu}+(1-a)}{Z_a(\theta)}
            \frac{\Gamma(b+\theta)\Gamma(1-b+\theta)}{\Gamma(1+2\theta)}\times\\
            & \left(-a {}_1F_1(b+\theta, 1+2\theta, (2a-1)\nu) + \frac{1-b+\theta}{1+2\theta}
            {}_1F_1(b+\theta, 2+2\theta, (2a-1)\nu) \right)
        \end{split}
    \]
    
    In the following figure we plot both the continuous and the discrete neutral model with and without hitchhiking (parameters: $\mu=0.025$, $\nu = 2$):

    \includegraphics[width=\textwidth]{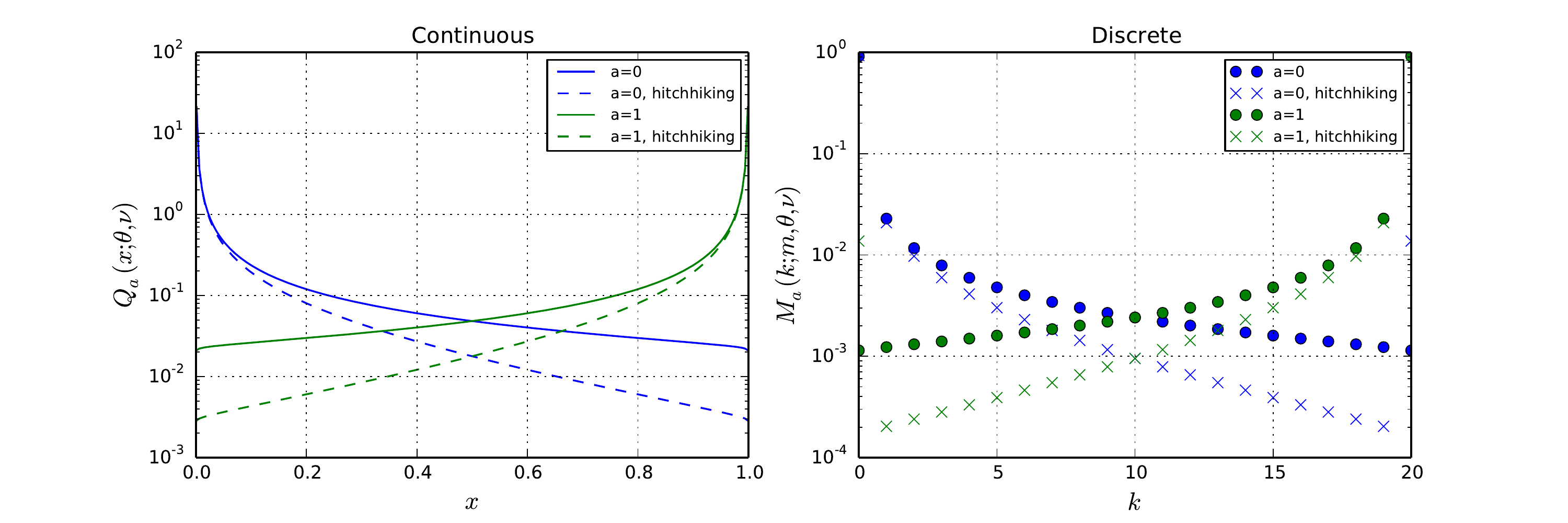}

    As expected, the hitchhiking model predicts fewer common variants in comparison to the standard model.

    \subsection{Selection}
    We can add selection to the standard model without hitchhiking, following \cite{Mustonen:2007im}:
    \[
        Q_a(x;\theta,\sigma)=\frac{1}{Z_a(\theta,\sigma)}(x(1-x))^{-1+\theta}\left(1-e^{\sigma((1-a)(x-1)+ax)}\right)
    \]
    with the normalization factor
    \[
        Z_a(\theta,\sigma)=\frac{\Gamma(\theta)^2}{\Gamma(2\theta)}
    \left(1-e^{-(1-a)\sigma}{}_1{}F_1(\theta,2\theta,\sigma)\right).
    \]
    
    The binomial moments are derived again by integration, similar to equation \ref{eq_binom_sampling}:
    \[
        \begin{split}
            M_a(k;m, \theta, \sigma) = & \binom{m}{k}\frac{1}{Z_a(\theta, \sigma)}
            \frac{\Gamma(k+\theta)\Gamma(-k+m+\theta)}{\Gamma(m+2\theta)}\times\\
            & \left(
                1 - e^{-(1-a)\sigma}{}_1F_1(k+\theta, m+2\theta, \sigma)
            \right)
        \end{split}
    \]
    
    We can now apply the same modification using an exponential factor for hitchhiking as we did in equation \ref{eq_qdist_hh}, which for the equilibrium distributions under selection leads to:
    \[
        Q(x; \mu, \sigma, \nu) = Q_a(x; \theta, \sigma)e^{-\nu((1-a)x+a(1-x))} + 
        \sum_{b=\{0,1\}}\delta(x-b)h_a(b,\theta,\sigma, \nu)
    \]
    with the hitchhiking fraction
    \[
        \begin{split}
        h_a(b, \theta, \sigma, \nu) = & M_a(b;1,\theta,\sigma) - \frac{1}{Z_a(\theta,\sigma)}
        \frac{\Gamma(b+\theta) \Gamma(-b + 1 + \theta)}{\Gamma(1+2\theta)} e^{-(1-a)\sigma-a\nu}\times\\
        & \left(
        e^{(1-a)\sigma} {}_1F_1(b+\theta, 1+2\theta, -(1-a)\nu + a\nu)
        -{}_1F_1(b+\theta, 1+2\theta, s-(1-a)\nu + a\nu)
        \right)
        \end{split}
    \]
    and the binomial moments
    \[
        \begin{split}
            M_a(k; m, \theta, \sigma, \nu) = & \binom{m}{k}\frac{1}{Z_a(\theta,\sigma)}
            \frac{\Gamma(k+\theta)\Gamma(-k+m+\theta)}{\Gamma(m + 2\theta)}e^{-(1-a)\sigma-a\nu}\times\\
            & \left(
            e^{(1-a)\sigma}{}_1F_1(k+\theta, m+2\theta, -(1-a)\nu + a\nu) - 
            {}_1F_1(k+\theta, m+2\theta, \sigma - (1-a)\nu + a\nu)
            \right)\\
            & + \sum_{b=\{0,1\}}\delta_{k,bm}h_a(b,\theta,\sigma, \nu)
        \end{split}
    \]

    We plot again both the continuous and the discrete neutral model with and without hitchhiking, but including selection (same parameters as above but $\sigma=2$)

    \includegraphics[width=\textwidth]{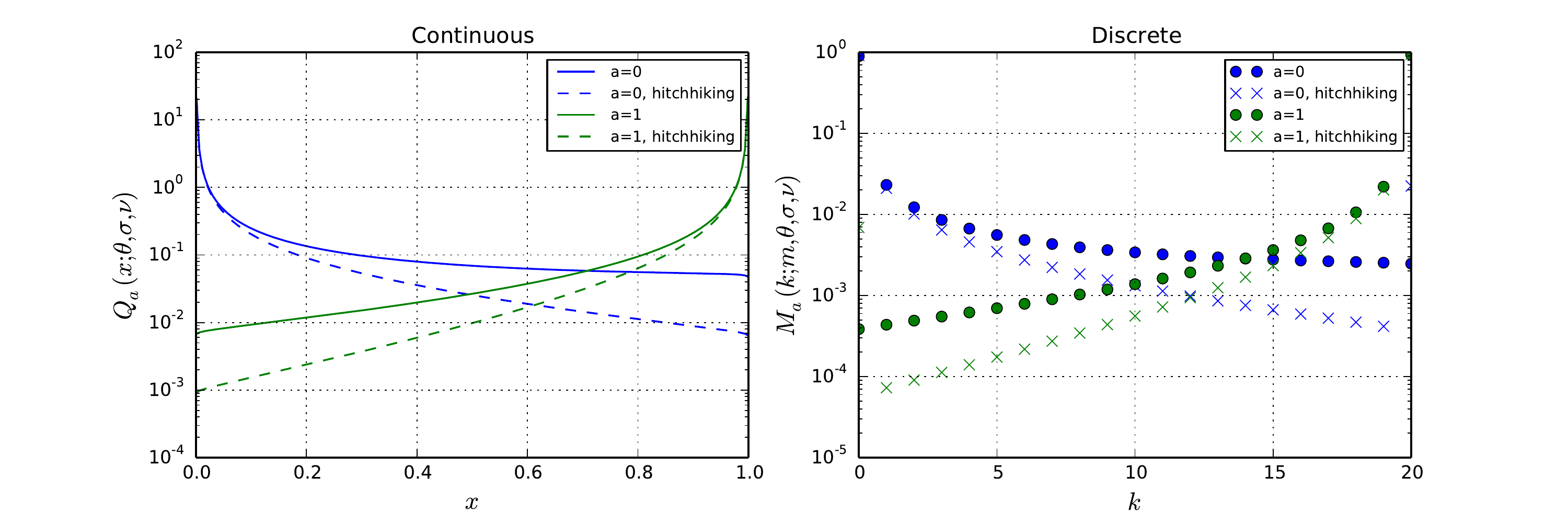}

    As expected, due to directional selection, allele frequencies are skewed towards the $B$-allele (i.e. higher $x$).
    
    \subsection{Substitutions and divergence from outgroup}
    \label{sec_subst}
    To model the divergence from a related species, we exploit the fact that we generally consider relatively small mutation rates, which leads to a separation of the substitution (fixation) time scale from the polymorphism time scale \cite{Lassig:2007iq}. In this regime, which corresponds to $\theta\ll1$, we can model fixations independently from polymorphisms as a Poisson process. Without hitchhiking and in the weak mutation regime, the rate per scaled unit time of this process is approximately \cite{Mustonen:2007im}
    \[
        u(\theta,\sigma) = \frac{\theta\sigma}{1-e^{-\sigma}}.
    \]
    
    Under hitchhiking with an effective rate $\nu$ (see above), we have previously shown \cite{Schiffels:2011fu} that the substitution rate changes to approximately
    \[
        u(\theta,\sigma,\nu)=\begin{cases}
            u(\theta,\sigma){}_2F_1\left(1, \frac{\nu}{\sigma}, 1+\frac{\nu}{\sigma}, 1-\frac{u(\theta,\sigma)}{\theta}\right) & \text{for  } \sigma>0\\
            \frac{\theta}{|\sigma|+\nu}\left(\frac{u(\theta,\sigma)}{\theta} |\sigma| + \nu\right) & \text{for  }\sigma<0,
        \end{cases}
    \]
    which effectively reduces the rate of beneficial mutations, and increases the rate of deleterious mutations, making them \"more neutral\".
    
    We abbreviate $u_+=u(\theta,+\sigma,\nu)$ and $u_-=u(\theta,-\sigma,\nu)$ and will omit the dependencies on $\theta$, $\sigma$ and $\nu$, which are always implicit in the following expressions. The rates $u_+$ and $u_-$ form a two state Markov process with states $a=\{0,1\}$ and a rate matrix
    \begin{equation}
        \label{eq_rateM}
        \mathbf{R}=\bordermatrix{
              &  0    &  1   \cr
            0 & -u_+  &  u_- \cr
            1 &  u_+  & -u_- 
         }
    \end{equation}
    and transition probability
    \[
    \mathbf{T}(\tau) = \exp\left(\mathbf{R}\,\tau\right) =
    \frac{1}{u_-+u_+}\left(\begin{matrix}
        u_- + u_+e^{-\tau(u_++u_-)} & u_- - u_-e^{-\tau(u_++u_-)} \cr
        u_+ - u_+e^{-\tau(u_++u_-)} & u_+ + u_-e^{-\tau(u_++u_-)}
    \end{matrix}\right).
    \]
    
    The equilibrium state $\boldsymbol{\lambda}$ for this transition probability is
    \[
        \boldsymbol{\Lambda} = \frac{1}{u_+ + u_-} \left(\begin{matrix} u_+\\u_-\end{matrix}\right)
    \]
    where we again left out the dependency on $\theta$, $\sigma$ and $\nu$ for brevity.
    
    This transition probability lets us write the probability to observe a frequency $k_1$ out of $m_1$ samples with allele $B$ in species 1, and $k_2$ out of $m_2$ samples in species 2, where both species share a common ancestral species at time $\tau$ in the past:
    
    \[
    M(k_1, k_2; m_1, m_2, \tau) = \sum_{a=0}^1\sum_{a_1=0}^1\sum_{a_2=0}^1
            (\boldsymbol{\Lambda})_a (\mathbf{T}(\tau))_{a_1,a} (\mathbf{T}(\tau))_{a_2,a} M_{a_1}(k_1;m_1) M_{a_2}(k_2;m_2).
    \]
    The indices in the matrix $\mathbf{T}$ and the vector $\boldsymbol{\Lambda}$ denote row and column, respectively. This expression makes use of the fact that the polymorphisms dynamics (reflected by $M_a(k;m)$) are independent of the substitution dynamics (reflected by $\mathbf{T}(\tau)$) in the weak mutation regime set by $\theta\ll1$.
    
    The outgroup-directed allele frequency as used in the data analysis is now simply a sum over the two cases in which the single outgroup-sample carries either of the two alleles:
    \[
        P(k; m, \tau, \theta, \sigma, \nu) = \sum_{k'=0}^1 M(k, k'; m, 1, \tau, \theta, \sigma, \nu).
    \]
    
    This is the most general allele frequency probability distribution, on which all models for parameter estimation as described in Material and Methods are based on. The following figure shows this probability for $\theta=0.025$ and $\tau=5$ and different values for $\sigma$ and $\nu$ as indicated:
    
    \begin{center}
        \includegraphics[width=0.6\textwidth]{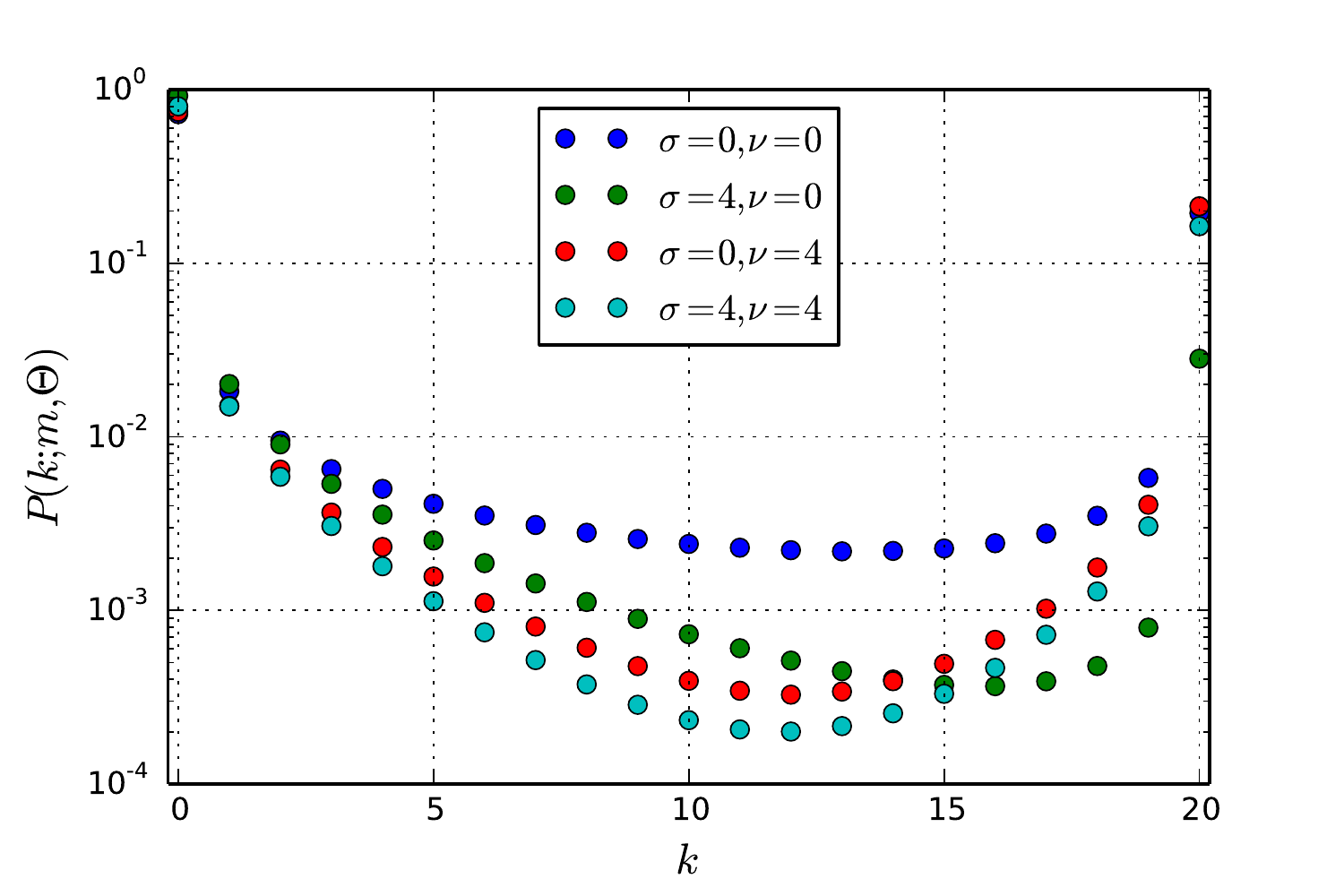}
    \end{center}
    
    Note that this probability is independent with respect to the sign of $\sigma$. The reason is that we compute the difference in the two species, without direction of an ancestral vs. derived allele. If $\sigma$ is negative, it means that the $A$ allele is the fitter one, but the probability to observe $k$ out of $m$ samples with a different allele than the outgroup is the same if $B$ was the fitter allele. We therefore treat $\sigma$ as a parameter on the positive domain of real values.
    
    In all of the above we have considered $N=N_0$, i.e. $\lambda=1$ (see section \ref{sec_def}). We can generalize by scaling all scaled parameters by $\lambda$:
    \[
        P(k; m, \tau, \theta, \sigma, \nu, \lambda) = \sum_{k'=0}^1 M(k, k'; m, 1, \tau/\lambda, \lambda\theta, \lambda\sigma, \lambda\nu).
    \]
    
    \section{Models}
    \subsection{Basic Models for synonymous sites}
    To work with real data and to infer parameters reliably, we define simplified subsets of the full model, by setting some parameters to default values. In particular, we define these basic models:

    \textbf{Unlinked adaptation model: }The unlinked model has as free parameters only the scaled mutation rate $\theta$ and the divergence time $\tau$. Other parameters are fixed, so that the resulting probability can be written as:
    \[
        P_\mathrm{unlinked}(k;m,\tau,\theta)=P(k;m,\tau,\theta,0,0,1).
    \]

    \textbf{Background selection model: }The background selection model (BGS) has as additional free parameter the effective population size $\lambda$:
    \[
        P_\mathrm{BGS}(k;m,\tau,\theta,\lambda)=P(k;m,\tau,\theta,0,0,\lambda)
    \]

    \textbf{Linked Adaptation model: }With hitchhiking, we use one further parameter $\nu$:
    \[
        P_\mathrm{linked}(k;m,\tau,\theta,\nu,\lambda)=P(k;m,\tau,\theta,0,\nu,\lambda)
    \]

    \textbf{Directional selection model: }We also define a model with background selection and direct selection (see Supplementary Figures S2 and S3):
    \[
        P_\mathrm{sel}(k;m,\tau,\theta,\sigma,\lambda)=P(k;m,\tau,\theta,\sigma,0,\lambda)
    \]
    
    \subsection{Mixed model for heterogeneous data sets}
    For our mixed model we add together these components with different weights:
    \begin{itemize}
        \item Neutral component: A fraction $c_n$ of sites evolves neutrally, but generally under hitchhiking.
        \item Weakly selected component: A fraction $c_w$ of sites evolves under weak directional selection.
        \item Adaptive component: At a fraction $c_a$ of sites we assume that adaptive evolution has generated fixed differences between the two species. This fraction accounts for an observed surplus of substitutions with respect to the neutral expectation.
        \item Conserved component: The remainder of the above, with fraction $c_c=1-c_n-c_w-c_a$ is assumed to be under strong directional selection which accounts for an observed surplus of conserved sites with respect to the neutral expectation.
    \end{itemize}

    Each component has its own specific outgroup-directed allele frequency distribution. First, the neutral component is simply one of the above derived basic models without selection:
    \begin{equation}
        \label{eq_Pn}
        P_n(k;m,\tau,\theta,\nu, \lambda) =
        \begin{cases}
            P_\mathrm{unlinked}(k; m,\tau,\theta)\\
            P_\mathrm{BGS}(k; m,\tau,\theta, \lambda)\\
            P_\mathrm{linked}(k; m,\tau,\theta, \nu, \lambda),
        \end{cases}
    \end{equation}
    defining three separate mixed models. The weakly selected component uses the full probability derived above, including a selection coefficient $\sigma$, which we typically constrain to $1<\sigma<150$:
    \begin{equation}
        \label{eq_Pw}
        P_w = P(k;m,\tau,\theta,\sigma,\nu,\lambda).
    \end{equation}
    The adaptive component is only a surplus of fixed differences between the two species, so it is simply
    \begin{equation}
        \label{eq_Pa}
        P_a(k;m) = \delta_{k,m}
    \end{equation}
    with the Kronecker-Delta which sets this probability to zero everywhere except at $k=m$ where it is one. Analogously, the conserved component is:
    \[
        P_c(k;m) = \delta_{k,0}
    \]

    Note that each of these components is normalized across $0\le k\le m$.

    The full probability of the mixed model is then:
    \[
        P(k;m,\Theta) = c_n P_n(k;m,\tau,\theta,\nu, \lambda) + c_w P_n(k;m,\tau,\theta,\sigma,\nu, \lambda) + c_a P_a(k;m) + (1-c_n-c_w-c_a) P_c(k;m).
    \]

    This full model has 7 parameters $\Theta=\{\tau, \theta, \sigma, \nu, c_n, c_w, c_a\}$.

    \subsection{Maximum Likelihood estimation}

    We consider a data set of outgroup-directed allele frequencies with a fixed sample size $m$. We denote the number of sites with allele frequency $k$ by $n_k$. The total likelihood of the data given parameters $\Theta$ is then:
    \[
        \mathcal{L}({n_k};m,\Theta) = \prod_{k=0}^m P(k;m,\Theta)^{n_k}.
    \]

    In practice we use the log-Likelihood
    \[
        \log\mathcal{L}({n_k};m,\Theta) = \sum_{k=0}^m n_k \log P(k;m,\Theta).
    \]

    The parameters are then learned by maximization of the log-Likelihood:
    \[
        \hat{\Theta} = \mathrm{argmax}_{\Theta'} \log\mathcal{L}({n_k};m,\Theta').
    \]

    As pointed out in the text, we typically follow a hierarchical protocol to estimate parameters from data. Assuming, all sites have been binned according to the local recombination rate (see Methods in the main article), we first use the unlinked model to infer $\tau$ (under free variation of $\theta$) from synonymous sites in the highest recombination bin. We then use the background selection and linked adaptation models to infer $\theta$, $\lambda$ and $\nu$ for each bin, keeping $\tau$ fixed at the value inferred from the highest recombination bin using the unlinked model. We then learn the rest of the parameters from non-neutral annotation categories using the mixed models, keeping the neutral parameters fixed to their values obtained from the background selection or the linked adaptation model. Numerical maximization is implemented using Powell's method \cite{Press}.
    
    \subsection{Types of substitutions in the mixed model}
    In the mixed model, we implemented different components which contribute differently to the amount of fixed differences between species. We make use of the three components $P_n$, $P_w$ and $P_a$ as defined in equations \ref{eq_Pn}, \ref{eq_Pw}, \ref{eq_Pa}. First, we can estimate the fraction of sites with neutral substitutions that have been fixed by drift alone:
    \[
        f_\mathrm{drift} =c_n P_n(k=m;m,\tau,\theta,\nu=0,\lambda)e^{-\nu},
    \]
    where $e^{-\nu}$ is the probability that no linked sweep occurs during the typical time of fixation of a neutral variant ($2N_0$ generations).
    We also define the fraction of sites with neutral substitutions fixed by drift and by hitchhiking:
    \[
        f_\mathrm{drift+HH} =c_n P_n(k=m;m,\tau,\theta,\nu,\lambda).
    \]
    From these two, we obtain the hitchhiking fraction via:
    \[
        f_\mathrm{HH}=f_\mathrm{drift+HH} -f_\mathrm{drift}.
    \]

    We obtain the fraction of sites with weakly selected substitutions that have been fixed by drift via:
    \[
        f_\mathrm{sel,drift} =c_w P_w(k=m;m,\tau,\theta,\sigma,\nu=0,\lambda)e^{-\nu},
    \]
    and the fraction of sites with weakly selected substitutions fixed by drift and hitchhiking
    \[
        f_\mathrm{sel,drift+HH} =c_w P(k=m;m,\tau,\theta,\sigma,\nu,\lambda),
    \]
    which allows us to derive the fraction of sites with weakly selected substitutions fixed by deleterious hitchhiking:
    \[
        f_\mathrm{sel,HH} =f_\mathrm{sel,drift+HH} -f_\mathrm{sel,drift}.
    \]
    We can also write down the fraction of sites with adaptive substitutions:
    \[
        f_\mathrm{adaptive} = c_a.
    \]

    \section{Fitness Flux}
    Fitness flux was introduced in \cite{Mustonen:2007im} as the product of the rate of substitutions and their average selection coefficient. To estimate fitness flux, we first compute the rate of adaptive substitutions per scaled unit time, $k_a$, from our fraction $f_a=c_a$ above:
    \[
        k_a = \frac{c_a}{2\tau},
    \]
    because $2\tau$ is the total branch length between the two species, and $c_a$ is the expected number of substitutions per site. If we knew the average selection coefficient of adaptive substitutions, $s_a$, the fitness flux $\Phi$ would be
    \begin{equation}
        \label{eq_Phiplus}
        \Phi = k_a s_a,
    \end{equation}
    per scaled unit time. We cannot measure $s_a$ directly from our framework, but we have an indirect measure using the rate of linked sweeps, $\nu$. As has been shown in \cite{Smith:1974vn} and \cite{Kaplan:1989tm}, the typical ``effect width'' of a single selective sweep with selection coefficient $s_a$ is
    \[
        w = \alpha \frac{s_a}{r},
    \]
    where $r$ is the recombination rate, and $\alpha$ is a constant, which depends on model assumptions and parameters such as the absolute effective population size. For relevant parameters, $\alpha$ lies between 0.1 and 0.5, as computed in \cite{Kaplan:1989tm}. Here we fix $\alpha=0.1$ to be conservative in estimating fitness flux and $s_a$ as follows: We can now relate the effective rate of linked sweeps, $\nu$, which is simply to total rate of adaptive mutations within a window of width $w$, with the fitness flux (see also \cite{Mustonen:2007im}):
    \[
        \nu = w k_a \gtrsim 0.1 k_a \frac{s_a}{r} = 0.1 \frac{\Phi}{r},
    \]
    So the rate of linked sweeps $\nu$ is directly linked to the fitness flux per recombination map length. In case we do not observe any positive rate of linked drivers, we estimate a conservative upper bound on the fitness flux using the inequality $\nu<1$ (with $\nu=1$ being a typical value that we can measure with confidence). In summary, we compute fitness flux as:
    \[
        \Phi \begin{cases}
            \lesssim 10 \nu r & \text{if $\nu>1$}\\
            \lesssim 10 r & \text{if $\nu\leq 1$}
        \end{cases},
    \]
    as an upper bound on fitness flux in regions with $r>0$. Having estimated $\Phi$, we can simply solve for $s_a$ using equation \ref{eq_Phiplus}.
    
    Similarly to the rate of beneficial mutations, we can estimate the total rate of weakly deleterious substitutions via
    \[
        k_d = \frac{f_\mathrm{sel,drift+HH}}{2\tau},
    \]
    which then lets us estimate a negative component to the fitness flux as
    \[
        \Phi_- = \frac{k_d}{2}\frac{\sigma}{2N_0},
    \]
    where $\sigma$ is the scaled selection coefficient as used in the mixed model, and the factor $1/2$ accounts for the fact that in equilibrium only half of the substitutions are deleterious, the other half is compensatory and hence beneficial. We used a diploid population size of $1.78\times 10^6$, as used in \cite{Sella:2009hs} and \cite{Andolfatto:2007cq}. We show this estimate in Supplementary Figure 6c). In the definitions above, fitness flux is defined in units of $1/2N_0$. However, it is intuitive to use as units $\mu/2N_0$, with $\mu$ estimated from our estimates of $\theta$ and the above population size. In these units, a fitness flux of $1$ can for example be generated by mutations with selection coefficients of $1/2N$, which fix with the neutral substitution rate ($\mu$). We use these units in Figure 5.
    
    To report a fitness flux per million years, we assume a generation time of 0.1 years in Drosophila, which yields a neutral substitution rate of one substitution per $0.1/\mu\approx 30\times10^6$ years.
    
    We also translate these rates to fitness flux per gene, for which we use an average number of 1,000 nonsynonymous sites per gene in the autosomes in Drosophila, obtained from the annotations described in methods.
    
    \section{Genetic Load}
    Genetic load as used here quantifies the amount of fitness loss generated by fixed deleterious mutations. The probability for a weakly selected site to be in the less fit state follows from the equilibrium state of the Markov process defined in equation \ref{eq_rateM}:
    \[
        \lambda_- = \frac{u_-}{u_+ + u_-},
    \]
    where the fixation rates $u_+$ and $u_-$ are described in section \ref{sec_subst}. The genetic load is then simply:
    \[
        l = \sigma c_w \lambda_-
    \]
    where $c_w$ is the fraction of weakly selected sites, and $\sigma$ is their selection coefficient, as defined in the mixed model.

\end{document}